\documentclass{jfm}
\usepackage{graphicx}
\usepackage{epstopdf, epsfig}
\usepackage{dcolumn}
\usepackage{bm}
\usepackage{amsfonts}
\usepackage{amssymb}
\usepackage{amsmath}    
\usepackage{graphicx}   
\usepackage{verbatim}   
\usepackage{color}      
\usepackage[colorlinks=true,linkcolor=blue,citecolor=blue,allcolors=blue]{hyperref}

\shorttitle{Suspension Dynamics of Droplets in Acoustic and Gravitational Fields}
\shortauthor{Jeyapradhap, Anjan and Karthick}

\title{Suspension Dynamics of Droplets in Acoustic and Gravitational Fields}

\author{Jeyapradhap Thirisangu\aff{1},
  Anjan Mahapatra\aff{1},
 \and Karthick Subramani\aff{1}
 \corresp{\email{karthick@iiitdm.ac.in}}}

\affiliation{\aff{1}Department of Mechanical Engineering, Indian Institute of Information Technology, Design and Manufacturing, Kancheepuram, Chennai-600127, India}

\begin{document}

\maketitle

\begin{abstract}

In the field of acoustic suspension or levitation of droplets against gravity, the application of Gorkov's acoustic radiation force for small particles (within the Rayleigh limit, i.e., $d \ll \lambda$) or its extensions to larger ones (beyond the Rayleigh limit, i.e., $d\gtrsim  \lambda$) is limited to predicting the suspension position of the droplet. Since this approach treats the droplet as a rigid particle, it fails to capture the fluid dynamics of the droplet and is also unsuitable for studying interfacial phenomena such as droplet deformation, splitting, or coalescence. In this work, we employ our recently developed acoustic body force in Eulerian form \citep{Thirisangu2024Aug}, which models the droplet as a fluid, to theoretically investigate the suspension dynamics of droplet subjected to standing waves through the interaction between acoustic, interfacial, and gravitational forces. Our theory predicts that when interfacial forces are dominant, the presence of positive and negative acoustic force regions within droplets exceeding the Rayleigh limit reduces the net acoustic force counteracting gravity. As a result, the suspension dynamics become highly dependent on droplet size, in contrast to droplets within the Rayleigh limit, where the dynamics remain size-independent. Thus, beyond the Rayleigh limit, as the droplet size to wavelength ratio ($d/\lambda$) increases, the critical acoustic energy density ($E_{cr}$) required to suspend the droplet initially rises sharply, which agrees with recent experimental results \citep{Thirisangu2023Dec}. After $E_{cr}$ reaches a local maximum at $d/\lambda \approx 0.65$, $E_{cr}$ exhibits a pattern of alternating decreases and increases, with each successive peak surpassing the previous one. Remarkably, our study reveals a size-dependent shifting of the suspension position between nodes and antinodes for droplets beyond the Rayleigh limit, whereas droplets within this limit maintain a consistent suspension position regardless of size. Furthermore, our theory predicts the sorting of the droplets based on the critical energy method, which was proposed in recent experimental work \citep{Thirisangu2023Dec}. 
Unlike the case where interfacial force dominates, when this force weakens, acoustic body force can deform the droplet. If the droplet contains regions with opposing acoustic forces, these forces compete with each other, potentially causing the droplet to split. If gravitational force is not sufficiently weak, droplet splitting becomes unequal or asymmetric.

\end{abstract}

\section{Introduction}\label{Sec 1}
The manipulation of matter through acoustic fields originates from the pioneering works of \cite{Chladni1787} and \cite{Michael1831Dec}, who explored particle patterns on vibrating plates. Later, \cite{kundt1874longitudinal} developed Kundt's tube, a device that used fine powders to visualize acoustic standing waves, offering an effective tool for studying sound propagation and resonance. \cite{king1934acoustic} formulated a theoretical expression for the radiation pressure on small rigid spheres, without considering the compressibility of the spheres. \cite{yosioka1955acoustic} later extended this theory for compressible spheres in line with experimental observations. \cite{gor1962forces} further developed this approach by formulating the acoustic radiation force ($\boldsymbol{F}_{ac}$) as the gradient of a potential, now widely known as the Gor'kov potential, and predominantly applied to particles within the Rayleigh limit (the droplet size ($d$) is much smaller than the wavelength ($\lambda$) of the acoustic wave ($d << \lambda$)).

\begin {equation}
\label{Eq 1}
    \boldsymbol{F}_{ac} = 4\pi d^3kE_{ac}\mathrm{sin}\mathrm{}(2ky)\frac{1}{3}\left(\frac{5\ \widetilde{\rho }-2}{2\widetilde{\rho}+1}-\frac{1}{\ \widetilde{\rho }\ \widetilde{c}^2}\right)\boldsymbol e_y .  
\end{equation}
\par Later, the above acoustic radiation force was employed to manipulate droplets and bubbles against gravity as follows. \cite{eller1968force} experimentally demonstrated the trapping of air bubbles in a liquid medium using acoustic standing waves, overcoming buoyancy forces. Following this, \cite{Crum1971Jul} successfully suspended small liquid droplets in the water against gravity, showing that the minimum pressure amplitude required for suspension is independent of droplet size. Subsequent works \citep{Hasegawa1979Janexp, Hasegawa1979Jantheory, Marston2017Sep, Baasch2018Jan, ospina2022particle, Rueckner2023Aug} have refined and extended the formula ($\boldsymbol{F}_{ac}$) to apply to particles beyond the Rayleigh limit ($d\gtrsim  \lambda$).

In most of the theoretical studies, droplets are treated as rigid particles, both within \citep{Crum1971Jul,  luo2016experimental, luo2017suspension}
and beyond the Rayleigh limit \citep{ Marston2017Sep, Baasch2018Jan, Yarin1998Feb, Zang2017Mar}. Consequently, the force ($\boldsymbol{F}_{ac}$) acting on droplets in the Lagrangian form, which is not fully compatible with the Navier-Stokes equation, fails to capture the fluid dynamics of droplets and is also unsuitable for studying interfacial phenomena such as droplet deformation, splitting, and coalescence. The above limitation is especially notable in theoretical studies of acoustic suspension/levitation against gravity, where the Gor'kov acoustic force ($\boldsymbol{F}_{ac}$) determines the suspension position of the droplet \citep{Lee1991Nov, Lee1994Nov, Yarin1998Feb, Tian1993Jun, Shi1996Apr, Di2018Oct, Naka2020Dec, Zang2017Mar}, but $\boldsymbol{F}_{ac}$ is not applied to describe deformation/splitting. Instead, deformation is typically evaluated by substituting acoustic radiation pressure into the Young-Laplace equation, which yields only static deformation shapes and fails to capture the dynamic, time-dependent behavior of droplets. A theoretical framework addressing the splitting of levitated droplets has not been developed so far \citep{Stone1989Jan, Danilov1992Nov, Foresti2013Jul,Hasegawa2019Nov, Naka2020Dec, Aoki2020May}.

In our previous work \citep{Thirisangu2024Aug}, we established a comprehensive theoretical framework to investigate the behaviour of droplets subjected to acoustic fields. Using the acoustic body force in Eulerian forms, the above framework governed the migration, deformation, and splitting of droplets. The prior study primarily concentrated on the interplay between acoustic and interfacial forces, omitting the influence of gravitational effects. In this study, we extend our theoretical framework to study the dynamics of droplet suspension in a liquid medium under acoustic fields, incorporating the combined effects of acoustic forces, interfacial forces, and gravity. We reveal that suspension dynamics of larger droplets become highly size-dependent, in contrast to droplets within the Rayleigh limit \citep{Crum1971Jul}, due to the presence of opposing acoustic forces within the droplet, which reduces the net acoustic force counteracting gravity. We also show that this size-dependent behaviour leads to phenomena such as size-dependent droplet switching between nodes and antinodes consistent with the rigid particle suspension/levitation dynamics works by \cite{ospina2022particle, Rueckner2023Aug}), alternating patterns in critical energy density, and droplet sorting based on critical energy density. Along with the above suspension position dynamics, this framework allows the study of the interfacial effects such as deformation, splitting, and coalescence during droplet suspension.

\section{Physics of the problem}\label{Sec 2}

When the immiscible fluid droplets are subjected to  gravity and acoustic fields, their dynamics are governed by incompressible mass continuity, momentum conservation, and  advection equations\citep{Landau1987Aug}, which are given below,

\begin{subequations}
\label{Eq 2}
\begin{equation}
\label{Eq 2a}
    \boldsymbol{\nabla} \cdot\textbf{\emph{v}}=0,
\end{equation} 
\begin{equation}
\begin{split}
\label{Eq 2b}
    \rho[\partial_t \textbf{\emph{v}} + (\textbf{\emph{v}}\cdot\boldsymbol{\nabla})\textbf{\emph{v}}] &= -\boldsymbol{\nabla} p +\eta\boldsymbol{\nabla}^2 \textbf{\emph{v}}   +\textbf{\emph{f}}_{g}+\textbf{\emph{f}}_{\sigma}+\textbf{\emph{f}}_{a},
\end{split}
\end{equation}
\begin{equation}
\label{Eq 2c}
   \partial_t \phi + \textbf{\emph{v}} \cdot \boldsymbol{\nabla} \phi = 0,
\end{equation}
\end{subequations}
where $\rho$ is the density, $\textbf{\emph{v}}$ is the velocity, $p$ is the pressure, $\eta$ is the dynamic viscosity of the fluid, ${\phi}$ is the phase fraction ($\phi=0$ for fluid $1$ and $\phi=1$ for fluid $2$), and $\textbf{\emph{f}}_{g}=-\rho g e_y$ is the gravity force. The interfacial force ($\textbf{\emph{f}}_{\sigma}$) is expressed as $\sigma \kappa \delta_{s} \textbf{n}$, where $\sigma$ is the surface tension, $\kappa$ is the curvature, $\delta_{s}$ is a surface Dirac delta function which is non-zero only on the interface, and $\textbf{n}$ is the unit normal. 

The slow-time phenomena ($v$,$p$), including acoustic radiation force, acoustic streaming, and the acoustic relocation of inhomogeneous fluids, are induced by the fast-time acoustic fields ($v_f$, $p_f$) resulting from the application of high-frequency acoustic waves. Neglecting the effects of acoustic streaming, the acoustic body force $\textbf{\emph{f}}_{a}$, as discussed in \cite{Rajendran2022Jun}, is expressed as:
\begin{equation}
\begin{split}
\label{Eq 3}   
   \textbf{\emph{f}}_{a}& = -\boldsymbol{\nabla}\cdot\langle\rho_s\textbf{\emph{v}}_f \otimes \textbf{\emph{v}}_f\rangle= \frac{1}{2}\boldsymbol{\nabla}(\kappa_0\langle|p_f|^2\rangle-\rho_0 \langle|\textbf{\emph{v}}_f|^2\rangle) \\&
    -\frac{1}{2}(\langle|p_f|^2\rangle\boldsymbol{\nabla}\kappa_0+\langle|\textbf{\emph{v}}_f|^2\rangle\boldsymbol{\nabla}\rho_0).   
\end{split}
\end{equation}.

Here, $\rho_o$ and $\kappa_o$ represent the background density and compressibility, respectively, while $p_f$ and $\textbf{\emph{v}}_f$ denote the fast-time-scale acoustic pressure and velocity fields induced by acoustic waves. The symbol $\langle...\rangle$ indicates time averaging over one oscillation period of the wave. For the standing acoustic wave applied along the y-direction, (pressure $p_f=p_a\sin(ky)$ and velocity $\textbf{\emph{v}}_f=\frac{p_a}{i\rho_0 c_0}\cos(ky) \boldsymbol e_y $) the above equation reduces to \citep{Rajendran2022Jun}, 


\begin{equation} 
\label{Eq 3.1}
\textbf{\emph{f}}_{a}=\boldsymbol{\nabla}\phi_{{a}}-E_{ac}\cos({2ky})\boldsymbol{\nabla}\hat{Z} \boldsymbol e_y  =\boldsymbol{\nabla}\phi_{{a}}+\textbf{\emph{f}}_{a1}.
\end{equation}

Where $E_{ac}=p_a^2/(4\rho_{avg}c_{avg}^2)$ is the acoustic energy density, $p_a$ is the pressure amplitude, $Z = \rho c$ denotes impedance, $\hat{Z}=Z/Z_{avg}$, $Z_{avg}=(Z_{1} +Z_{2})/2$,  $c$ speed of sound and $\boldsymbol e_y$ is the unit normal vector along the $y$ direction, where the subscript  ‘avg’ denotes the respective average quantities of droplet and continuous medium. It is important to note that Eq. \ref{Eq 3} is valid for the plane standing acoustic wave considered in this study. Whereas, for an acoustic beam or a travelling acoustic wave, the corresponding $p_f$ and $v_f$ are to be substituted in the Eq. \ref{Eq 3}. In the equation Eq. \ref{Eq 3.1}, the first term represents a conservative or gradient term that induces pressure within the bounded domain without causing fluid flow, and the second term is responsible for the relocation of inhomogeneous fluids.  Thus, the non-gradient body force term $(\textbf{\emph{f}}_{a1})$, in the above equation is responsible for droplet movement under acoustic fields. This term is further expanded as discussed in \cite{Thirisangu2024Aug} as follows, 

\begin{equation} 
\label{Eq 4}
\textbf{\emph{f}}_{a1}=-E_{ac}\cos({2ky})\boldsymbol{\nabla}\hat{Z} \boldsymbol e_y=-\boldsymbol{\nabla} \left(\cos({2kx})\hat{Z} \right)-\textbf{\emph{f}}_{ac}.
\end{equation}

The non-gradient body force term $(\textbf{\emph{f}}_{ac})$ becomes,

\begin{equation} 
\label{Eq 5}
\textbf{\emph{f}}_{ac}=-2kE_{ac}\sin({2ky})\hat{Z} \boldsymbol e_y.
\end{equation}

\cite{Thirisangu2024Aug} showed that both $\textbf{\emph{f}}_{a1}$ (Eq. \ref{Eq 4})
and $\textbf{\emph{f}}_{ac}$ (Eq. \ref{Eq 5}) yields the same result, except the variation in the background pressure due to the existence of the gradient term in the Eq. \ref{Eq 4}. The advantage of $\textbf{\emph{f}}_{a1}$ (Eq. \ref{Eq 5}) over $\textbf{\emph{f}}_{ac}$ (Eq. \ref{Eq 4}) is that the structure of $\textbf{\emph{f}}_{ac}$ is consistent with Gorkov's force Equation ($\boldsymbol{F}_{ac}$) for small particles, which enables better scaling and allows straightforward prediction of droplet behaviour.

\begin{figure}
    \center
    \includegraphics[width=0.9\linewidth]{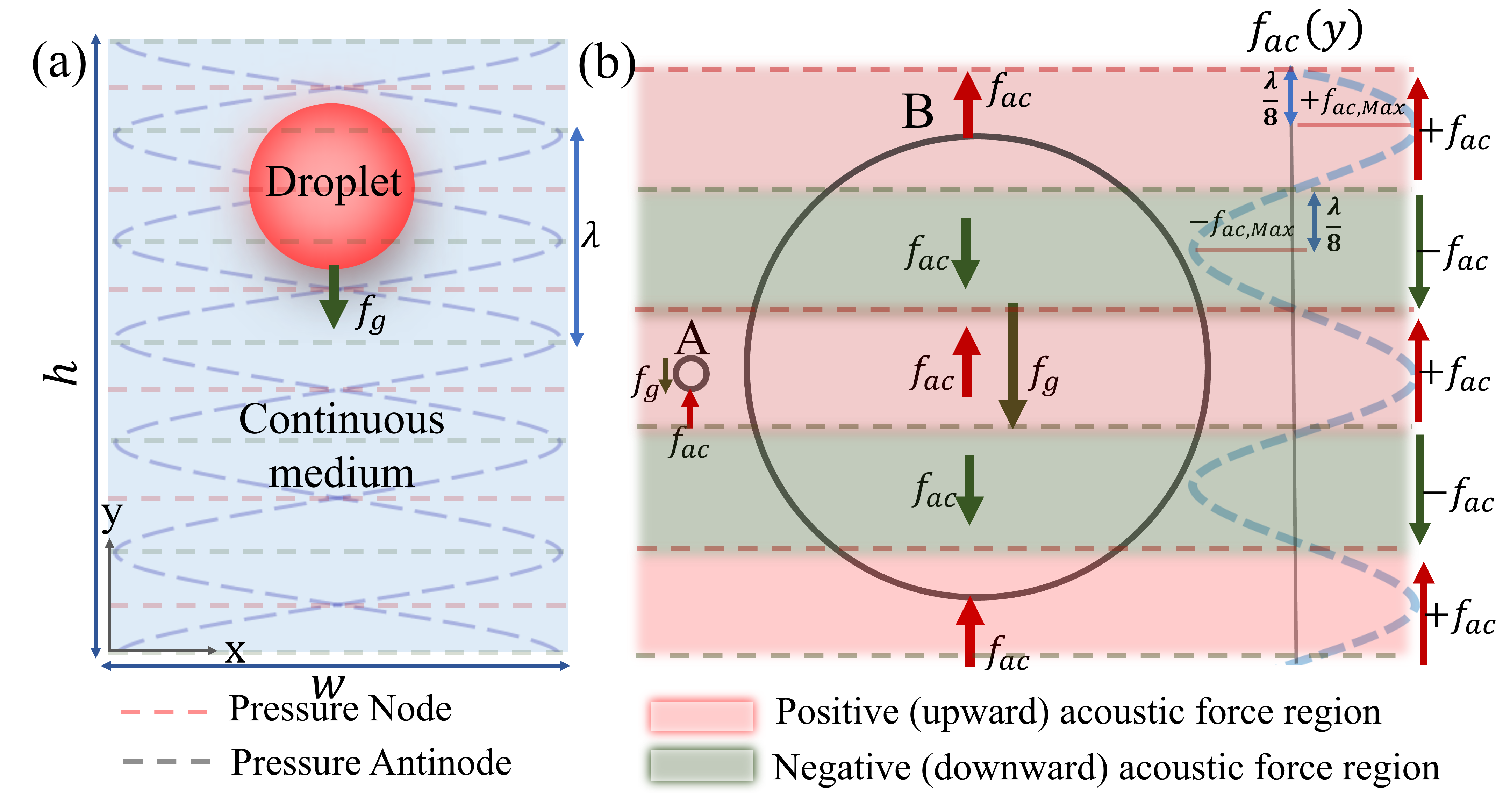}        
    \caption{\label{physics} (a) Schematic representation of the droplet suspended in a continuous liquid medium subjected to acoustic standing wave with the $\lambda \ll h$. (b) Schematic representation of acoustic force distribution within the smaller droplet (A) and the larger droplet (B) of a size beyond the Rayleigh limit ($d \gtrsim \lambda$).}
\end{figure} 

Using Gorkov's equation (Eq. \ref{Eq 1}), \cite{Crum1971Jul} demonstrated that for droplets within the Rayleigh limit, the minimum pressure amplitude ($p_a$) or acoustic energy density ($E_{ac}$) required to suspend the droplet is independent of its size, as both the acoustic and gravitational forces scale proportionally with the droplet volume ($\sim a^3$). However, for droplets beyond the Rayleigh limit, our recent experimental work \citep{Thirisangu2023Dec} showed that the minimum $p_a$ or $E_{ac}$ required to suspend the droplets increases with the droplet size. In this work, we now proceed to show that the $\textbf{\emph{f}}_{ac}$ (Eq.\ref{Eq 5}) in Eulerian form effectively predicts this behaviour by illustrating how larger droplets encompass both positive and negative force regions, as shown in Fig. \ref{physics}(b). The partial cancellation of these opposing forces reduces the net acoustic force acting against gravity, making suspension strongly size-dependent. We also show that this size dependence gives rise to other intriguing phenomena, such as the size-dependent switching of droplets between a pressure node to a pressure antinode, sorting of the droplets, and unequal splitting of droplets. To enhance readability, 'pressure node' and 'pressure antinode' are referred to as 'node' and 'antinode,' respectively, throughout the manuscript.

\section{Numerical method}\label{Sec 3} 

The numerical analysis of droplets subjected to acoustic fields is performed by simultaneously solving the governing equations: the continuity equation (Eq. \ref{Eq 2a}), the advection-diffusion equation (Eq. \ref{Eq 2c}), and the momentum equation (Eq. \ref{Eq 2b}) coupled with the acoustic force equation (Eq. \ref{Eq 5}) as the body force term, using COMSOL Multiphysics 6.0. The effects of gravity and interfacial forces are also included in the computational framework. In the simulations, the droplet interface is modelled as diffusive to prevent singularities and improve numerical stability. This is achieved by incorporating a diffusion term into Eq. \ref{Eq 2c}, expressed as follows: $\partial_t\phi_s + \textbf{\emph{v}}_s\cdot\boldsymbol{\nabla}\phi_s=\gamma\boldsymbol{\nabla}\cdot\left( 
\epsilon\boldsymbol{\nabla}\phi_s - \phi_s(1-\phi_s)\frac{\boldsymbol{\nabla \phi_s}}{\left| \boldsymbol{\nabla \phi_s} \right|} \right)$.The parameter $\epsilon$ defines the width of the transition region where $\phi_s$ smoothly varies from 0 to 1, ensuring a gradual interface transition. The parameter $\gamma$ controls the degree of reinitialization or stabilization of the level set function, enhancing numerical stability. A two-dimensional channel of width ($w$) $6$  mm and height ($h$) $8$ mm containing a droplet in a continuous domain is considered for the study. An acoustic wave with a wavelength $'\lambda'$ is applied along the height of the channel in the $y$ direction with the no-slip boundary condition enforced at all walls. Since this is a closed domain, a pressure constraint is imposed at a single point to stabilize the pressure field. The number of grids used in this study is $8,67,560$, with the maximum size of the grid being 12.53 microns, beyond which the flow fields are unaffected by the increase in the number of grids. The following fluid combinations are considered in this study: water droplet in mineral oil, silicon droplet in mineral oil, silicon droplet in olive oil, and water droplet in olive oil (Fig. \ref{Critical}). However, to qualitatively compare the two-dimensional numerical results with our recent experimental findings \citep{Thirisangu2023Dec}, all the results presented in this study are for water droplets (${\rho}_{\mathrm{water}} = 1000$ $kg/m^3$, $c_{\mathrm{water}} = 1480$ $m/s$, and $\mu_{\mathrm{water}} = 0.8$ $mPa.s$) in mineral oil (${\rho}_{\mathrm{oil}} = 857.5$ $kg/m^3$, $c_{\mathrm{oil}} = 1440$ $m/s$, and $\mu_{\mathrm{oil}} = 26.5$ $mPa.s$), with an applied wavelength of $\lambda = 2$ mm. The properties of all these fluids are taken from \cite{hemachandran2019relocation}.

\section{Results and discussion}\label{Sec 4} 
When a higher-density immiscible fluid droplet is introduced into a lower-density continuous medium in the absence of an acoustic field, the droplet undergoes uniform downward motion due to the balance between gravity and drag forces. However, the introduction of the acoustic field disrupts this uniform motion, enabling control over the droplet's movement. This section presents a detailed investigation of the droplet dynamics under the combined effects of acoustic and gravitational forces. From Sections \ref{Section 4.1} to \ref{Section 4.5}, the droplet dynamics are studied when the interfacial effects dominate over the acoustic and gravitational forces, making droplet deformation negligible. As the interfacial force weakens, the acoustic force begins to significantly influence the shape of the droplet, which is discussed in Section \ref{Section 4.5}. For brevity, the velocity of the droplet in the absence of acoustic force (under the balance between gravitational and drag forces) is referred to as 'uniform velocity' throughout the manuscript.

\subsection{Suspension characteristics of the droplet}\label{Section 4.1}
\begin{figure}
    \center
    \includegraphics[width=0.8\linewidth]{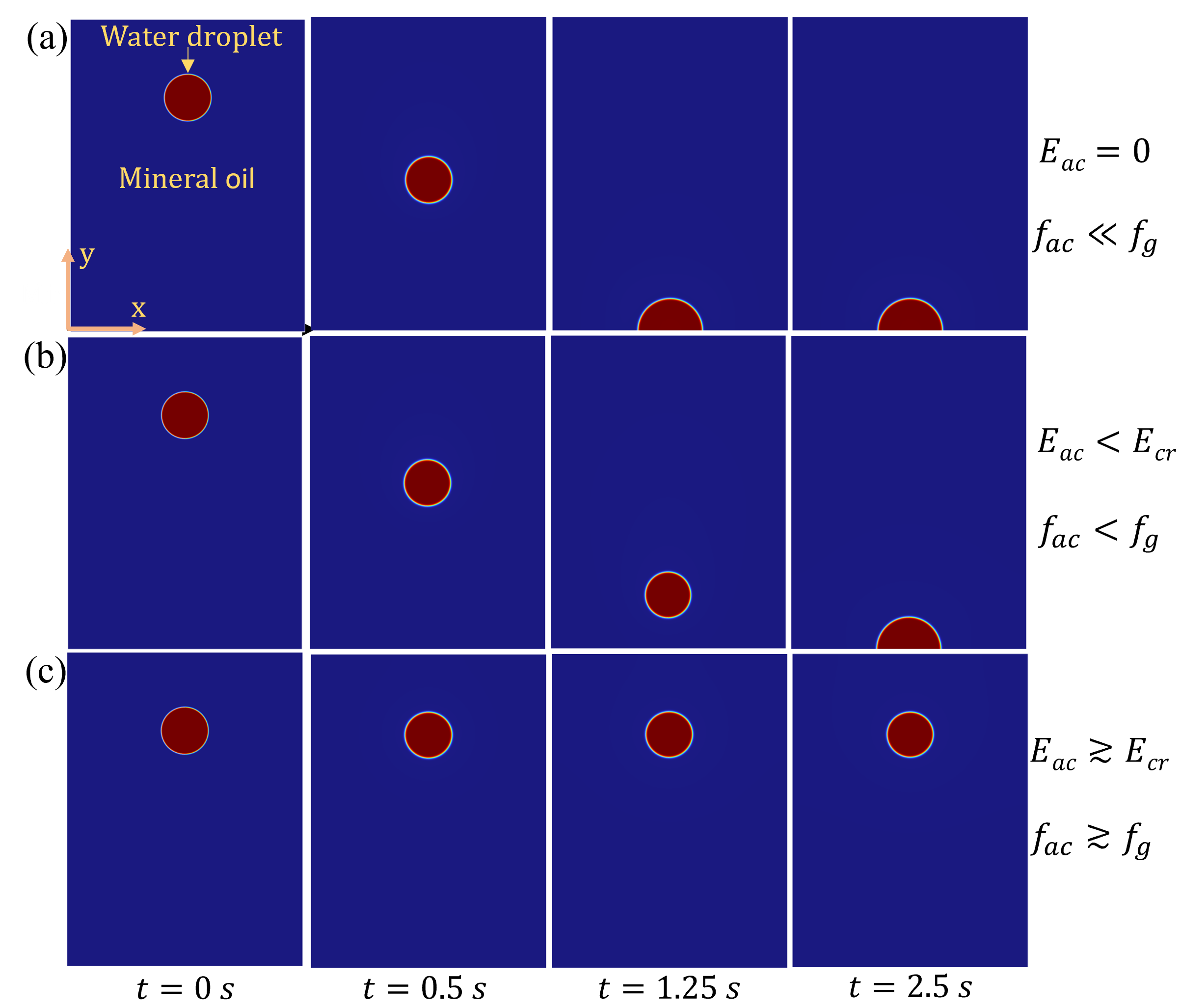}
    \caption{\label{Suspending} 
    Suspension characteristics of water droplet ($d = 1.5 \, \text{mm}$) in mineral oil subjected to varying acoustic energy densities: (a) The droplet settles with uniform motion in the absence of acoustic energy density (${E_{ac}} = 0$); (b) The droplet settles with delayed settling time at an acoustic energy density below the critical value (${E_{ac}} = 8.3 \, \text{J/m}^3$); (c) The droplet is suspended at an acoustic energy density above the critical value (${E_{ac}} = 11.3  \,  \text{J/m}^3$).}
\end{figure}

Figure \ref{Suspending} shows the behaviour of a water droplet ($d=$ 1.2 mm) in mineral oil subjected to the acoustic field (Eq. \ref{Eq 5}) of varying acoustic energy density ($E_{ac}$). If the applied acoustic energy density overcomes gravity ($E_{ac} \gtrsim E_{cr}$), then the droplet suspends as shown in Fig. \ref{Suspending}(c). If the applied acoustic energy density is not enough to overcome gravity ($E_{ac} < E_{cr}$), then the uniform motion of the droplet as seen in the absence of acoustic force (Fig. \ref{Suspending}(a)) is disturbed and the settling time of the droplet is delayed (Fig. \ref{Suspending}(b)). This change in the settling time is due to the non-uniform acoustic force distribution along the $y$ direction, as shown in Fig. \ref{physics}(b). Since the wavelength ($\lambda=c_{0}/f=2$ mm) of the acoustic fields (in Eq. \ref{Eq 5}) is much lesser than the height ($H=8$ mm) of the fluid domain ($\lambda\ll H $), a series of nodes and anti-nodes are produced along the $y$ direction. During the settling process, when the droplet passes through this series of nodes and antinodes, the velocity of the droplet increses relative to the uniform velocity in the negative acoustic force region (shaded green colour in Fig. \ref{physics}(b)), where $\textbf{\emph{f}}_{ac}$ acts downward along with $\textbf{\emph{f}}_{g}$. Conversely, the velocity of the droplet decreases relative to the uniform velocity in the positive acoustic force region (shaded red colour in Fig. \ref{physics}(b)), where the acoustic force ($\textbf{\emph{f}}_{ac}$) acts upward against gravity. Upon applying enough acoustic energy density ($E_{cr}=$ 11.3 $\text{J/m}^3$ for $d=$ 1.2 mm), the droplet suspends near the node as the acoustic force overcomes gravitational force (Fig. \ref{physics}(c)). Remarkably, the suspension characteristics of the water droplets in mineral oil, as shown in Fig \ref{Suspending}, qualitatively agree with our recent experimental results \citep{Thirisangu2023Dec}. In the following section, the critical acoustic energy density ($E_{cr}$) required to suspend the droplet of different sizes is discussed in detail.

\subsection{Critical acoustic energy density\label{Section 4.2}}

\begin{figure}
    \center
    \includegraphics[width=0.9\linewidth]{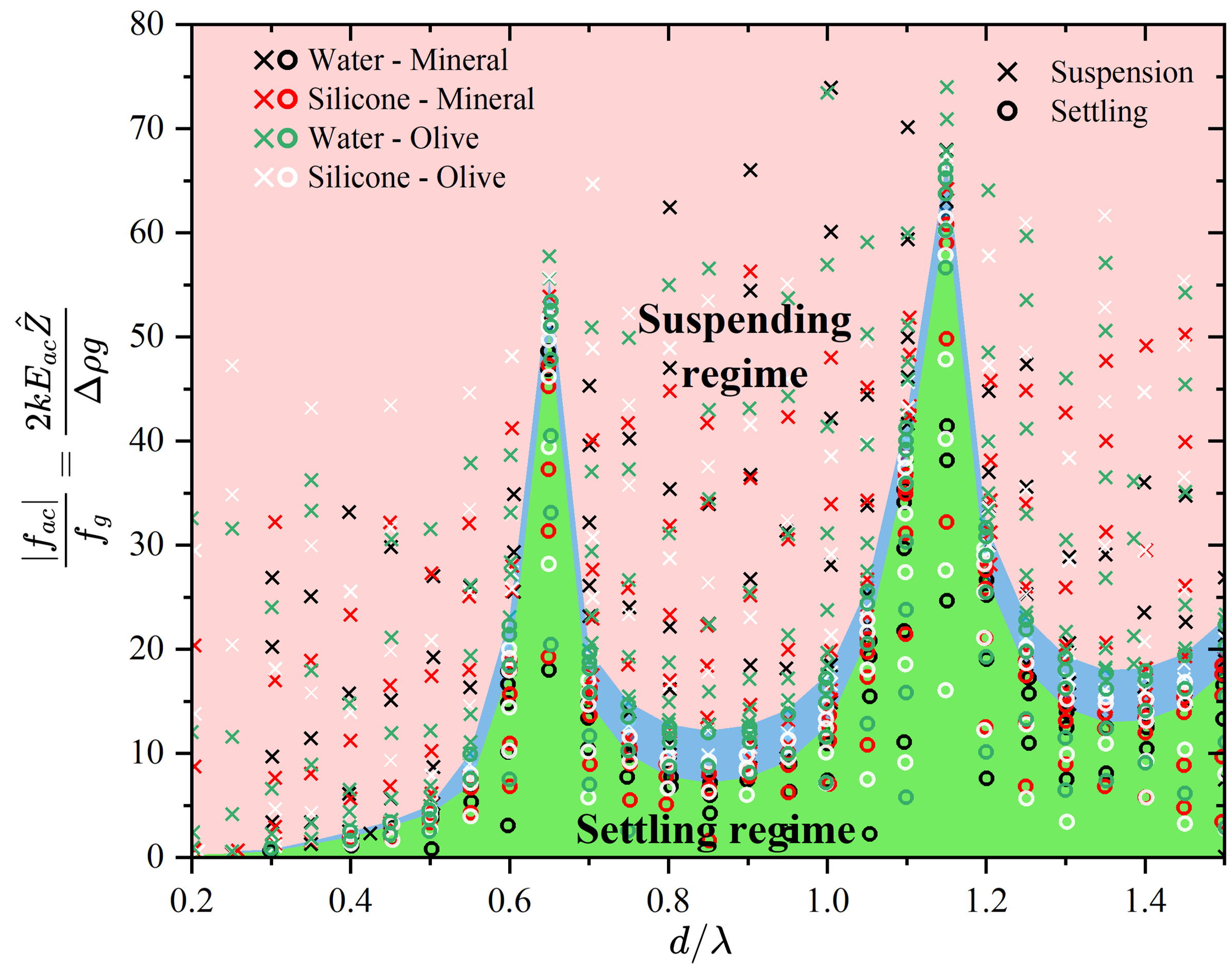}
    \caption{\label{Critical} 
     Suspending and Settling regimes of droplets beyond the Rayleigh limit ($d \gtrsim \lambda$) due to interplay between acoustic ($\textbf{\emph{f}}_{ac}$) and gravity ($\textbf{\emph{f}}_{g}$) forces. The suspending regime (pink) and settling regime (green) are separated by an intermediate regime (blue), in which some combinations of fluids suspend while others settle. For the given fluid combination, the first cross mark (just above the settling regime) represents the critical $\frac{\left| f_{ac} \right|}{f_{g}}$ required for droplet suspension (from which the $E_{cr}$ can be obtained). All the data points correspond to the simulation results. }
\end{figure}
When the droplets of varying sizes are subjected to acoustic fields, the critical acoustic energy density ($E_{cr}$) required to suspend the droplets also varies, as shown in Fig. \ref{Critical}(a). Within the Rayleigh limit ($a \ll \lambda$), $E_{cr}$ required for droplet suspension is independent of droplet size, as both the acoustic and gravitational forces scale with the droplet volume ($a^3$) \citep{ Crum1971Jul}. It is important to note that droplets of this size can be accommodated entirely within either the positive or negative acoustic force region (refer to droplet A in Fig. \ref{physics}(b)). If the droplet size increases beyond the Rayleigh limit, both positive and negative acoustic force regions exist within the droplet due to the presence of multiple nodes and antinodes (refer to droplet B in Fig. \ref{physics}(b)). The portion of the droplet located in the positive acoustic force region experiences an upward force ($\textbf{\emph{f}}_{ac}$ is positive), while the portion in the negative acoustic force region experiences a downward force ($\textbf{\emph{f}}_{ac}$ is negative). These opposing forces partially cancel out each other, resulting in a net acoustic force acting on the droplet that is significantly lower than the force that would act if either the positive or negative force were uniformly distributed across the same droplet. Consequently, as droplet size to wavelength ratio ($d/\lambda$) increases beyond the Rayleigh limit, $E_{cr}$ required to suspend the droplet also increases exponentially, reaching a peak at  $\sim 0.65$ $d/\lambda$, as shown in Fig. \ref{Critical}(a). This is due to the available volume for the net positive acoustic force decreasing up to $\sim 0.65$ $d/\lambda$. As a result, more energy ($E_{cr}$) is required to overcome gravitational force to suspend the droplet.
This theoretical result is consistent with recent experimental results \citep{Thirisangu2023Dec}.
Interestingly, beyond this peak, as the $d/\lambda$ further increases, the $E_{cr}$ required to suspend the droplet decreases up to the $\frac {d}{\lambda} \approx 0.85$ and then increases again, reaching the peak at $\sim 1.15$ $d/\lambda$, as shown in the Fig. \ref{Critical}(a). Remarkably, it is seen that the magnitude of the successive peak/trough is greater than that of the previous one (Fig. \ref{Critical}(a)). 
The reason for the sudden decrease in the $E_{cr}$ in the interval of $0.65 \lesssim d/\lambda \lesssim 0.85$ is that the increase in the available volume for the positive net acoustic force acting on the droplet. In contrast, in the interval $0.85 \lesssim d/\lambda \lesssim 1.15$, the available volume for the positive net acoustic force decreases, which demands more $E_{cr}$ to counteract gravity and suspend the droplet. 



\subsection{Shifting of the suspending position\label{Section 4.3}}

\begin{figure}
    \center
    \includegraphics[width=0.75\linewidth]{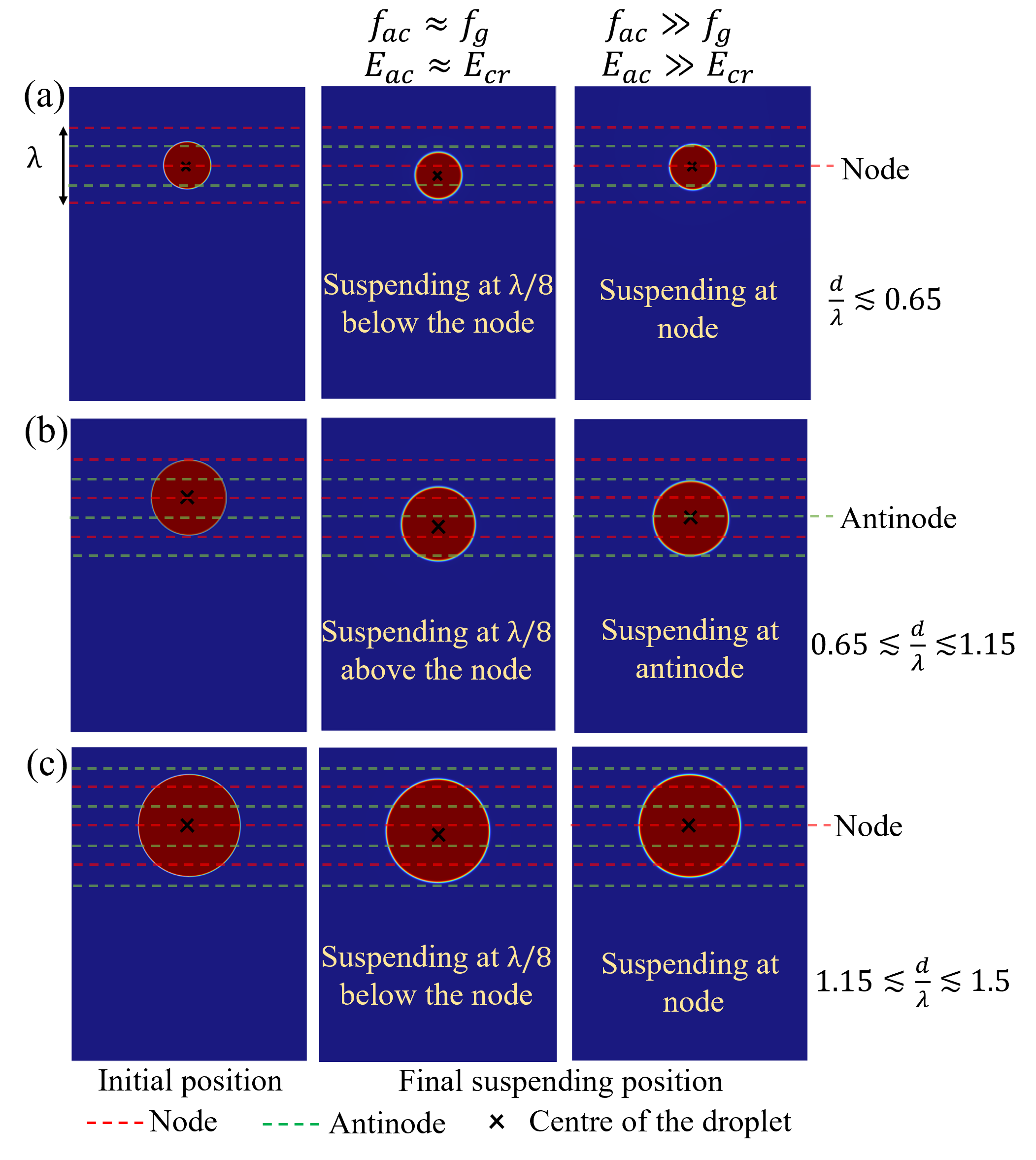}
    \caption{\label{Switching} Size dependent switching of the droplet between node and antinode: (a) Droplets with a size range $\frac{d}{\lambda} < 0.65$ are getting suspended near the nodes ($d = 1.2 \, \text{mm}$), (b) Droplets with a size range $0.65 < \frac{d}{\lambda} < 1.15$ are getting suspended near the antinodes ($d = 1.9 \, \text{mm}$), and (c) Droplets with a size range $1.15 < \frac{d}{\lambda} < 1.5$ are getting suspended near the nodes ($d = 2.6 \, \text{mm}$). (Note: In this study, all droplets are introduced at the node, and the results remain the same even when the droplets are introduced at the antinode.)}
\end{figure}

 Within the Rayleigh limit ($d \ll \lambda$), in the absence of gravitational force or under the condition where acoustic force dominates the gravitational force ($\textbf{\emph{f}}_{g}=0$ or ${f}_{ac} \gg {f}_{g}$), a water droplet (high acoustic impedance) in mineral oil (low acoustic impedance) medium migrate towards the nearest pressure node when subjected to acoustic fields. Whereas beyond the Rayleigh limit ($d \gtrsim \lambda$), under the same condition ($\textbf{\emph{f}}_{g}=0$ or ${f}_{ac} \gg {f}_{g}$), the migration of the droplet or the suspending position of the droplet becomes size-dependent as shown in Fig. \ref{Switching}. Upon applying the $E_{ac} \gg E_{cr}$ (${f}_{ac} \gg {f}_{g}$), for the droplet of sizes $\frac{d}{\lambda} \lesssim 0.65$ and $1.15 \lesssim d/\lambda \lesssim 1.5$, the suspension position of the centre of the droplet is seen to be suspended at the node as shown in Fig. \ref{Switching}(a) and (c). In contrast, the droplet of sizes $0.65 \lesssim d/\lambda \lesssim 1.15$ is seen to be suspended at the antinode (Fig. \ref{Switching}(b)). 
 
  \par In the presence of both gravity and acoustic force, at the critical acoustic energy density ($E_{ac} \approx E_{cr}$ or ${f}_{ac} \approx {f}_{g}$), the suspending position of the droplets within the Rayleigh limit ($d \ll \lambda$), is located at distance $\lambda/8$ below the pressure node \citep{Crum1971Jul}. This occurs due to the sinusoidal variation of the acoustic radiation force, which reaches its maximum upward magnitude at a distance $\lambda/8$ below the node (counteracting gravity) and its maximum downward magnitude at a distance of $\lambda/8$ above the node (aiding gravity) as shown in the Fig. \ref{physics}(b). Upon applying $E_{ac} \gg E_{cr}$ (${f}_{ac} \gg {f}_{g}$), the droplet suspending position shifts toward nodes. Whereas beyond the Rayleigh limit ($d \gtrsim \lambda$), for droplet with sizes $\frac{d}{\lambda} \lesssim 0.65$ and $1.15 \lesssim d/\lambda \lesssim 1.5$, at the critical acoustic energy density ($E_{cr}$ or ${f}_{ac} \approx {f}_{g}$), the suspending position of the centre of the droplet is seen to be located close to $\lambda/8$ below the pressure node as shown in Fig. \ref{Switching}(a) and (c). Due to the change in force direction (this result is consistent with the rigid particle behaviour of \cite{ ospina2022particle}), droplets with sizes $0.65 \lesssim d/\lambda \lesssim 1.15$ are seen to be suspended close to $\lambda/8$ above the pressure node for the applied critical acoustic energy density as shown in Fig. \ref{Switching}(b). The following section provides a detailed analysis of the influence of acoustic fields on droplet velocity and settling time (when $E_{ac} < E_{cr}$).

\subsection{Velocity and settling time\label{Section 4.4}}

\begin{figure}
\center
    \includegraphics[width=1\linewidth]{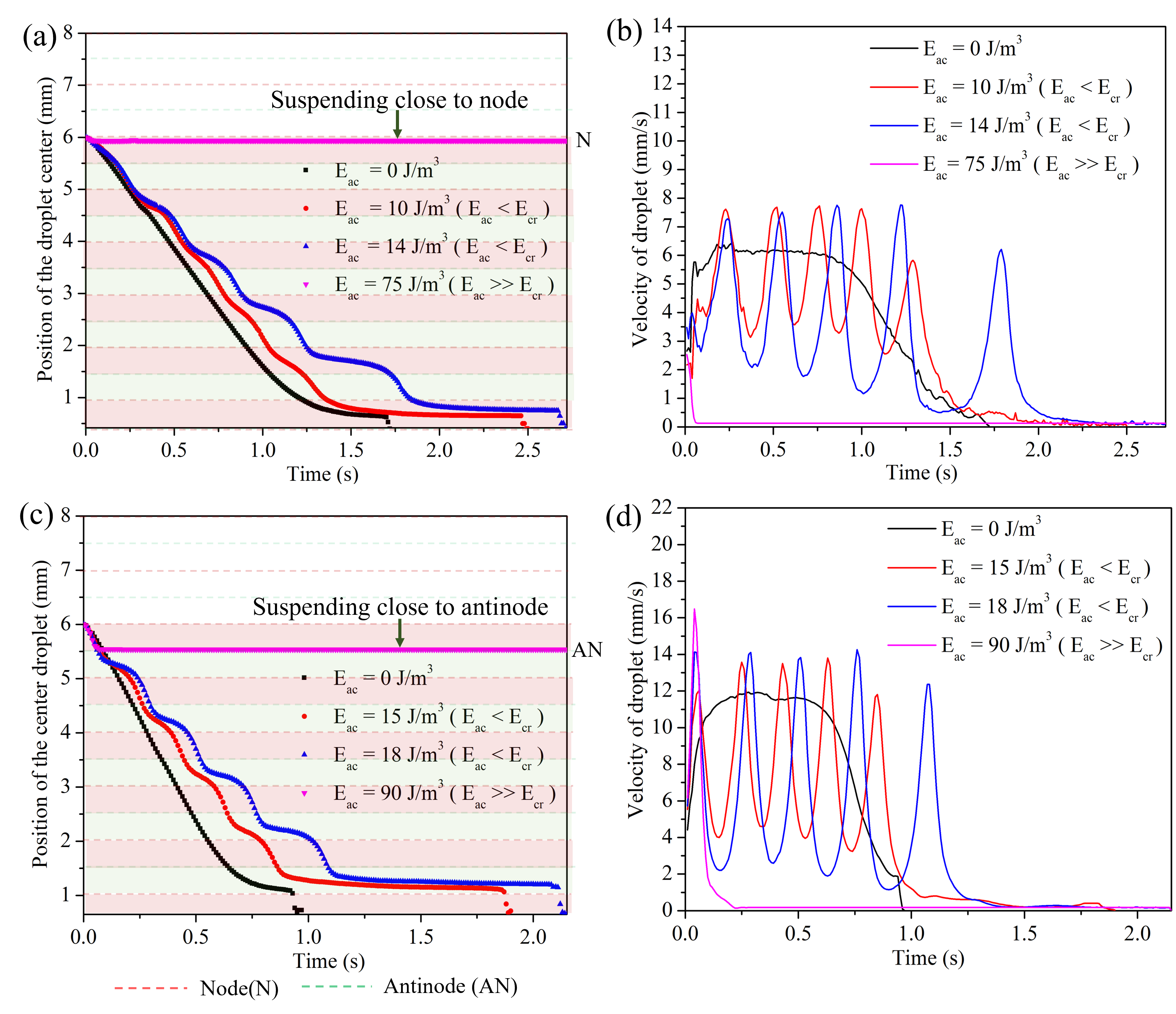}
    \caption{\label{velocity} For droplets with size ranges $\frac{d}{\lambda} \lesssim 0.65$ and $1.15 \lesssim \frac{d}{\lambda} \lesssim 1.5$, suspended at the node: (a) The variation of the location of the droplet center ($d = 1.2$ mm) with time and (b) the change in the velocity of the droplet with varying acoustic energy density. For droplets with a size range $0.65 \lesssim \frac{d}{\lambda} \lesssim 1.15$, suspended at the antinode: (c) The variation of the location of the droplet center ($d = 2.1$ mm) with time and (b) the change in the velocity of the droplet with varying acoustic energy density.}
\end{figure}
When the droplet is introduced at the node or antinode, the net acoustic force acting on it is zero, as the droplet is symmetrically positioned within both the positive and negative acoustic force regions. This symmetry causes the upward and downward forces to cancel each other out. As a result, gravity becomes the dominant force, causing the droplet to descend from its initial position. As the droplet begins to move between the node and antinode or vice versa, the net acoustic force becomes non-zero due to the asymmetry in the acoustic force regions within the droplet. If the net acoustic force becomes positive (acting upward against gravity) and upon applying the $E_{ac}\geq E_{cr}$, the droplet suspends, as shown in Fig. \ref{velocity}(a). However, if the net acoustic force is insufficient to counteract gravity ($E_{ac} < E_{cr}$), the droplet settles with delayed settling time compared to the settling time of the droplet without acoustics, as shown in Fig. \ref{velocity}(a). During this settling process, when the centre of the droplet is within the region where the net acoustic force acting on the droplet is positive (shaded red color in the Fig. \ref{velocity}(a)), its velocity decreases relative to the uniform velocity in the absence of acoustic forces. Conversely, when the centre of the droplet is within the region where the net acoustic force acting on the droplet is negative (shaded green colour in Fig. \ref{velocity}(a)), its velocity increases compared to the uniform velocity of the droplet in the case of without acoustics. Despite the alternative increases and decreases of velocity under acoustic fields, the delay in the settling time is due to the residence time of the droplet within each half-wavelength distance being more than the residence time of a droplet without acoustics. 

It is important to note that, for droplets within the size ranges $\frac{d}{\lambda} \lesssim 0.65$ and $1.15 \lesssim \frac{d}{\lambda} \lesssim 1.5$, the net acoustic force acting on the droplet is positive, when the center of the droplet is within quarter-wavelength below the node (shaded in red in Fig. \ref{velocity}(a)). Whereas, the net acoustic force acting on the droplet is negative, when the center of the droplet is within quarter-wavelength above the node (shaded in green in Fig. \ref{velocity}(a)). Consequently, droplets within this size range are seen to be suspended near the node (Fig. \ref{velocity}(a) and Fig. \ref{Switching}) with the applied $E_{ac}\gg E_{cr}$. In contrast to the above size, for droplets within the size range $0.65 \lesssim \frac{d}{\lambda} \lesssim 1.15$, the net acoustic force acting on the droplet is negative, when the center of the droplet is within quarter-wavelength below the node (shaded in red in Fig. \ref{velocity}(a)). Whereas, the net acoustic force acting on the droplet is positive, when the center of the droplet is within quarter-wavelength above the node (shaded in green in Fig. \ref{velocity}(a)). Consequently, droplets within this size range are seen to be suspended near the antinode (Fig. \ref{velocity}(a) and Fig. \ref{Switching}) with the applied $E_{ac}\gg E_{cr}$.




\subsection{Sorting of the droplet based on critical energy method\label{Section 4.5}}

\begin{figure}
\center
    \includegraphics[width=0.9\linewidth]{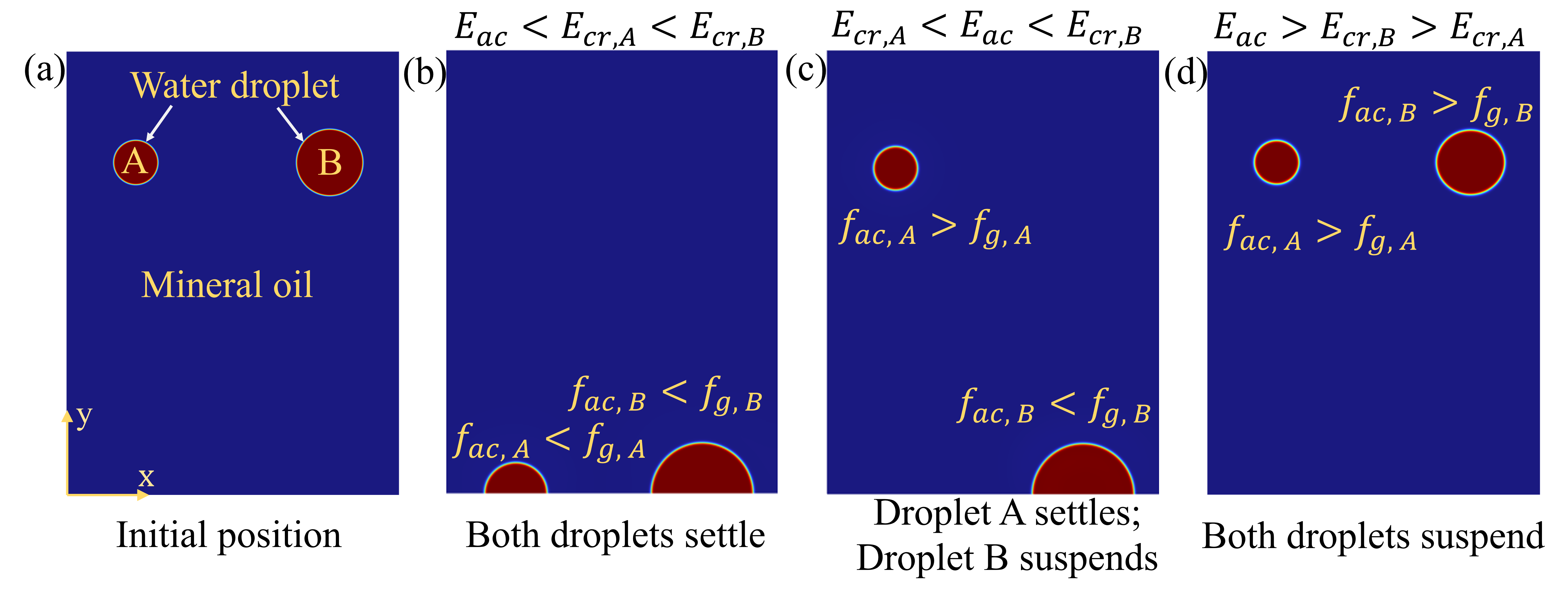}
    \caption{\label{sorting} Droplet sorting based on critical energy method. (a) The initial position of the droplet. (b) When the applied acoustic energy density ($E_{ac}$) is lower than the critical energy densities of both droplets ($E_{ac}<E_{cr,A}<E_{cr,B}$). (c) When the applied $E_{ac}$ is in between the critical energy densities of the droplet A and B ($E_{cr,A} < E_{ac} < E_{cr,B}$), the droplet A suspends while the droplet B settles. (d) When $E_{ac}$ exceeds the critical energy densities of both droplets ($E_{ac}>E_{cr,A}>E_{cr,B}$).}
\end{figure}
In our recent experimental work \citep{Thirisangu2023Dec}, the sorting of droplets of size beyond the Rayleigh limit based on the critical power/ critical energy method was demonstrated. We now proceed to show that our theoretical model predicts the experimental results for droplet sorting based on the critical energy method. Interestingly, for two droplets of different sizes beyond the Rayleigh limit, the critical acoustic energy density ($E_{cr}$) required to suspend each droplet varies: one droplet requires a higher $E_{cr}$, while the other requires a lower $E_{cr}$. When an acoustic energy density ($E_{ac}$) is applied between these two $E_{cr}$ values, only the droplet with the lower $E_{cr}$ suspends, while the droplet with the higher $E_{cr}$ settles. Thus, droplets can be effectively sorted based on the critical energy method. Figure \ref{sorting} illustrates this sorting technique clearly. When the applied acoustic energy density ($E_{ac}$) is lower than the critical energy densities of both droplets ($E_{ac}<E_{cr,A}<E_{cr,B}$), both droplets settle, as shown in Fig. \ref{sorting}(b). If $E_{cr,A} < E_{ac} < E_{cr,B}$, droplet A suspends while droplet B settles, as shown in Fig. \ref{sorting}(c). When $E_{ac}$ exceeds the critical energy densities of both droplets ($E_{ac}>E_{cr,A}>E_{cr,B}$), both droplets suspend, as depicted in Fig. \ref{sorting}(d).

The method described above significantly differs from two widely employed acoustic sorting techniques for small particles:  contrast factor-based sorting and size-based sorting. In contrast factor-based sorting \citep{petersson2005continuous}, particles with a positive contrast factor migrate toward the pressure node, while those with a negative contrast factor move toward the antinode.  In size-based sorting \citep{petersson2007free}, the acoustic radiation force scales with particle size, causing larger particles to migrate more quickly to the node or antinode than smaller particles. However, given sufficient time, particles of all sizes will ultimately converge at the node or antinode. In contrast to the size-based sorting method, our approach allows droplets of one size to remain suspended while droplets of another size settle indefinitely until the critical acoustic forces are applied to them.

\subsection{Interfacial effects\label{section 4.6}}

\begin{figure}
\center
    \includegraphics[width=1\linewidth]{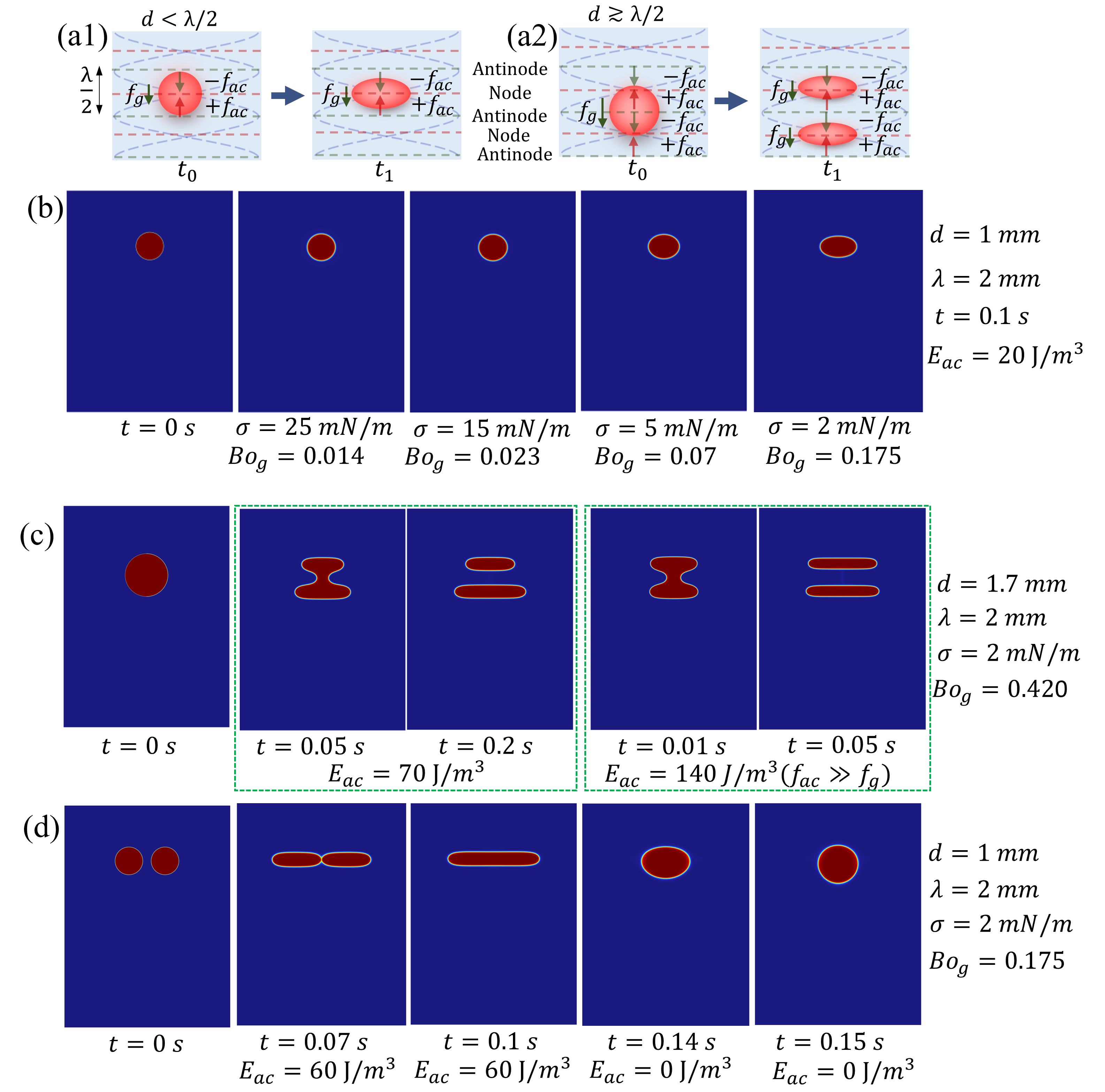}
    \caption{\label{surface} Effect of surface tension on droplet shape while suspending. (a1) Schematic representation of deformation of the droplet introduced at the node. (a2) Schematic representation of splitting of the droplet introduced at the antinode. (b) Droplet deformation: When droplets of size $d < \lambda/2$ are introduced at the node with $E_{ac} > E_{cr}$ causes it to suspend at the node and deform towards the nodes. As the surface tension coefficient ($\sigma$) decreases, the droplet deforms more. (c) Droplet splitting: When a droplet of size $d \gtrsim \lambda/2$ is introduced at the antinode with enough $E_{ac}$ to overcome gravity, it splits into two daughter droplets of different sizes (asymmetric splitting). When $E_{ac}$ is significantly dominating gravity, the droplet splits into two daughter droplets of the same size (symmetric splitting). (d) Droplet coalescence:  When droplets of size  $d < \lambda/2$ are introduced at a node adjacent to each other at the node, they deform towards the node and coalesce with acoustic energy density $E_{ac}>E_{cr}$.}
\end{figure}



Since the maximum Bond number $\left ( Bo = \Delta \rho g r^{2}/\sigma  \right )$ considered until the previous sections is 0.063 (for radius, $r = 1.5$ mm and $\sigma = 50$ mN/m), which is much less than 1, the dominant interfacial force over the acoustic and gravitational forces, enabling the droplet to retain its shape with negligible deformation.  We now proceed to present the influence of droplet shape when the interfacial force weakens (Fig. \ref{surface}).

Upon suspending the droplets of size $d < \lambda/2$ introduced at the node (Fig. \ref{surface}(a)) with the applied $E_{ac} > E_{cr}$, the positive acoustic force (the region below the node of the droplet centre) and the negative acoustic force (the region above the node of the droplet centre) are acting towards the centre of the droplet as shown in the Fig. \ref{surface}(a1)). This causes the droplet to squeeze/deform towards the node, as shown in Fig. \ref{surface}(b), particularly when the interfacial force is weak. The deformation of the droplet increases as the interfacial force weakens with a reduction in the surface tension coefficient ($\sigma$) under constant acoustic energy density ($E_{ac} > E_{cr}$), as shown in Fig. \ref{surface}(b).

When the interfacial force is weak ($\sigma = 2$ mN/m), upon suspending the droplets of size $d \gtrsim \lambda/2$ introduced at the antinode with the applied $E_{ac} \gg E_{cr}$, the positive acoustic force (the region above the antinode of the droplet centre) and the negative acoustic force (the region below the antinode of the droplet centre) are acting away from the centre of the droplet as shown in the Fig. \ref{surface}(a2). This causes the droplet to split into two daughter droplets, each moving to the nearest node as shown in Fig. \ref{surface}(c). When the applied acoustic energy density is just sufficient enough to overcome gravity for splitting, the gravitational effect causes 
asymmetrical splitting, where the lower daughter droplet is larger than the upper one, as shown in Fig. \ref{surface}(c). However, when the applied acoustic energy density significantly dominates the gravity (${f}_{ac} \gg {f}_{g}$), the daughter droplets are equal in size (symmetrical splitting), as shown in Fig. \ref{surface}(c). The deformation of the daughter droplet increases as the applied acoustic energy density ($E_{ac} \gg E_{cr}$) increases, as shown in Fig. \ref{surface}.

When droplets of size  $d < \lambda/2$ are introduced adjacent to each other at the node with the acoustic energy density $E_{ac} > E_{cr}$, the opposing forces within each droplet cause them to deform towards the nodes and coalesce, as shown in Fig. \ref{surface}(d). Once the acoustic field is turned off ($E_{ac}=0$ $J/m^3$), the coalesced droplet begins to settle.

\section{Conclusion}

This theoretical study examines the suspension dynamics of droplets under acoustic fields, by modeling droplet as fluid domain governed by the interplay of acoustic, interfacial, and gravitational forces. For droplets beyond the Rayleigh limit, we showed that their suspension dynamics become highly size-dependent, in contrast to droplets within the Rayleigh limit. This is due to the presence of regions of positive and negative acoustic forces within the droplet, which reduce the net acoustic force opposing gravity. We also highlighted that this size-dependent behavior leads to phenomena such as size dependent droplet switching between nodes and antinodes, alternating patterns in critical energy density, and droplet sorting based on critical energy density.


In addition to the significant insights and potential applications highlighted in this study, it is equally important to acknowledge its limitations. The droplets were modelled in two dimensions, resulting in only qualitative agreement with experimental observations that involve the three-dimensional behaviour of droplets. Consequently, deviations in the relationship between critical acoustic energy density ($E_{cr}$) and $d/\lambda$ ratio are expected. Furthermore, the steady $E_{ac}$ approach, which assumes first-order acoustic fields such as wave pressure and velocity, successfully captures the observed phenomena but does not account for position-dependent variations in $E_{ac}$. Although these variations are minimal due to the droplet's small volume relative to the fluid domain, addressing them would require solving first-order acoustic fields alongside slow timescale equations, a more complex and computationally intensive method. Additionally, this study is limited to liquid-liquid suspensions without extending to liquid-in-air levitation, which will be explored in future work.

\textbf{Funding}: This work was supported by the Department of Science $\&$ Technology—Science and Engineering Research Board (DST-SERB) via Grant No. SRG/2021/002180 and the Department of Science $\&$ Technology—Fund for Improvement of Science $\&$ Technology Infrastructure (DST-FIST) via Grant No. SR/FST/ET-I/2021/815.\\

\textbf{Declaration of Interests}: The authors report no conflict of interest.

\bibliographystyle{jfm}
\bibliography{jfm-instructions}

\begin{thebibliography}{36}
\expandafter\ifx\csname natexlab\endcsname\relax\def\natexlab#1{#1}\fi
\def\au#1{#1} \def\ed#1{#1} \def\yr#1{#1}\def\at#1{#1}\def\jt#1{\textit{#1}} \def\bt#1{#1}\def\bvol#1{\textbf{#1}} \def\vol#1{#1} \def\pg#1{#1} \def\publ#1{#1}\def\arxiv#1{#1}\def\org#1{#1}\def\st#1{\textit{#1}}

\bibitem[Aoki \& Hasegawa(2020)]{Aoki2020May}
{\sc \au{Aoki, K.} \& \au{Hasegawa, K.}} \yr{2020}  \at{{Acoustically induced breakup of levitated droplets}}.  \jt{AIP Adv.}  \bvol{10}~(5).

\bibitem[Baasch \& Dual(2018)]{Baasch2018Jan}
{\sc \au{Baasch, Thierry} \& \au{Dual, J{\ifmmode\ddot{u}\else\"{u}\fi}rg}} \yr{2018}  \at{{Acoustofluidic particle dynamics: Beyond the Rayleigh limit}}.  \jt{J. Acoust. Soc. Am.}  \bvol{143}~(1),  \pg{509--519}.

\bibitem[Chladni(1787)]{Chladni1787}
{\sc \au{Chladni, Ernst}} \yr{1787} {\em Entdeckungen uber die Theorie des Klanges\/}.  \publ{Leipzig, Germany: Weidmanns, Erben und Reich}.

\bibitem[Crum(1971)]{Crum1971Jul}
{\sc \au{Crum, Lawrence~A.}} \yr{1971}  \at{{Acoustic Force on a Liquid Droplet in an Acoustic Stationary Wave}}.  \jt{J. Acoust. Soc. Am.}  \bvol{50}~(1B),  \pg{157--163}.

\bibitem[Danilov \& Mironov(1992)]{Danilov1992Nov}
{\sc \au{Danilov, S.~D.} \& \au{Mironov, M.~A.}} \yr{1992}  \at{{Breakup of a droplet in a high{-}intensity sound field}}.  \jt{J. Acoust. Soc. Am.}  \bvol{92}~(5),  \pg{2747--2755}.

\bibitem[Di {\em et~al.\/}(2018)Di, Zhang, Li, Lin, Li, Li, Binks, Chen \& Zang]{Di2018Oct}
{\sc \au{Di, Wenli}, \au{Zhang, Zehui}, \au{Li, Lin}, \au{Lin, Kejun}, \au{Li, Jun}, \au{Li, Xiaoguang}, \au{Binks, Bernard~P.}, \au{Chen, Xiaopeng} \& \au{Zang, Duyang}} \yr{2018}  \at{{Shape evolution and bubble formation of acoustically levitated drops}}.  \jt{Phys. Rev. Fluids}  \bvol{3}~(10),  \pg{103606}.

\bibitem[Eller(1968)]{eller1968force}
{\sc \au{Eller, Anthony}} \yr{1968}  \at{{Force on a Bubble in a Standing Acoustic Wave}}.  \jt{J. Acoust. Soc. Am.}  \bvol{43}~(1),  \pg{170--171}.

\bibitem[Faraday(1831)]{Michael1831Dec}
{\sc \au{Faraday, Michael}} \yr{1831}  \at{{XVII. On a peculiar class of acoustical figures; and on certain forms assumed by groups of particles upon vibrating elastic surfaces}}.  \jt{Philos. Trans. R. Soc. Lond.}  \bvol{121},  \pg{299--340}.

\bibitem[Foresti {\em et~al.\/}(2013)Foresti, Nabavi, Klingauf, Ferrari \& Poulikakos]{Foresti2013Jul}
{\sc \au{Foresti, Daniele}, \au{Nabavi, Majid}, \au{Klingauf, Mirko}, \au{Ferrari, Aldo} \& \au{Poulikakos, Dimos}} \yr{2013}  \at{{Acoustophoretic contactless transport and handling of matter in air}}.  \jt{Proc. Natl. Acad. Sci. U.S.A.}  \bvol{110}~(31),  \pg{12549--12554}.

\bibitem[Gor'kov(1962)]{gor1962forces}
{\sc \au{Gor'kov, Lev~Petrovich}} \yr{1962} On the forces acting on a small particle in an acoustical field in an ideal fluid.  \bt{In {\em Sov. Phys.-Doklady\/}}, ,  \vol{vol.~6},  \pg{pp. 773--775}.

\bibitem[Hasegawa {\em et~al.\/}(2019)Hasegawa, Watanabe \& Abe]{Hasegawa2019Nov}
{\sc \au{Hasegawa, Koji}, \au{Watanabe, Ayumu} \& \au{Abe, Yutaka}} \yr{2019}  \at{{Acoustic Manipulation of Droplets under Reduced Gravity}}.  \jt{Sci. Rep.}  \bvol{9}~(16603),  \pg{1--8}.

\bibitem[Hasegawa(1979{\natexlab{{\em a\/}}})]{Hasegawa1979Janexp}
{\sc \au{Hasegawa, Takahi}} \yr{1979{\natexlab{{\em a\/}}}}  \at{{Acoustic radiation force on a sphere in a quasistationary wave field{\ifmmode---\else\textemdash\fi}experiment}}.  \jt{J. Acoust. Soc. Am.}  \bvol{65}~(1),  \pg{41--44}.

\bibitem[Hasegawa(1979{\natexlab{{\em b\/}}})]{Hasegawa1979Jantheory}
{\sc \au{Hasegawa, Takahi}} \yr{1979{\natexlab{{\em b\/}}}}  \at{{Acoustic radiation force on a sphere in a quasistationary wave field{\ifmmode---\else\textemdash\fi}theory}}.  \jt{J. Acoust. Soc. Am.}  \bvol{65}~(1),  \pg{32--40}.

\bibitem[Hemachandran {\em et~al.\/}(2019)Hemachandran, Karthick, Laurell \& Sen]{hemachandran2019relocation}
{\sc \au{Hemachandran, E}, \au{Karthick, S}, \au{Laurell, T} \& \au{Sen, AK}} \yr{2019}  \at{Relocation of coflowing immiscible liquids under acoustic field in a microchannel}.  \jt{Europhysics Letters}  \bvol{125}~(5),  \pg{54002}.

\bibitem[King(1934)]{king1934acoustic}
{\sc \au{King, King~Louis}} \yr{1934}  \at{{On the acoustic radiation pressure on spheres}}.  \jt{Proc. R. Soc. London A - Math. Phys. Sci.}  \bvol{147}~(861),  \pg{212--240}.

\bibitem[Kundt \& Lehmann(1874)]{kundt1874longitudinal}
{\sc \au{Kundt, A.} \& \au{Lehmann, O.}} \yr{1874}  \at{Longitudinal vibrations and acoustic figures in cylindrical columns of liquids}.  \jt{Ann. Phys.}  \bvol{229}~(9),  \pg{1--12}.

\bibitem[Landau \& Lifshitz(1987)]{Landau1987Aug}
{\sc \au{Landau, L.~D.} \& \au{Lifshitz, E.~M.}} \yr{1987} {\em {Fluid Mechanics}\/}.  \publ{Oxford, England, UK: Pergamon}.

\bibitem[Lee {\em et~al.\/}(1991)Lee, Anilkumar \& Wang]{Lee1991Nov}
{\sc \au{Lee, C.~P.}, \au{Anilkumar, A.~V.} \& \au{Wang, T.~G.}} \yr{1991}  \at{{Static shape and instability of an acoustically levitated liquid drop}}.  \jt{Phys. Fluids A}  \bvol{3}~(11),  \pg{2497--2515}.

\bibitem[Lee {\em et~al.\/}(1994)Lee, Anilkumar \& Wang]{Lee1994Nov}
{\sc \au{Lee, C.~P.}, \au{Anilkumar, A.~V.} \& \au{Wang, T.~G.}} \yr{1994}  \at{{Static shape of an acoustically levitated drop with wave{\textendash}drop interaction}}.  \jt{Phys. Fluids}  \bvol{6}~(11),  \pg{3554--3566}.

\bibitem[Luo {\em et~al.\/}(2017)Luo, Cao, Ren, Yan \& He]{luo2017suspension}
{\sc \au{Luo, Xiaoming}, \au{Cao, Juhang}, \au{Ren, Jing}, \au{Yan, Haipeng} \& \au{He, Limin}} \yr{2017}  \at{{Suspension characteristics of water droplet in oil under ultrasonic standing waves}}.  \jt{Ultrason. Sonochem.}  \bvol{39},  \pg{461--466}.

\bibitem[Luo {\em et~al.\/}(2016)Luo, He, Wang, Yan \& Qin]{luo2016experimental}
{\sc \au{Luo, Xiaoming}, \au{He, Limin}, \au{Wang, Hongping}, \au{Yan, Haipeng} \& \au{Qin, Yahua}} \yr{2016}  \at{{An experimental study on the motion of water droplets in oil under ultrasonic irradiation}}.  \jt{Ultrason. Sonochem.}  \bvol{28},  \pg{110--117}.

\bibitem[Marston(2017)]{Marston2017Sep}
{\sc \au{Marston, Philip~L.}} \yr{2017}  \at{{Finite-size radiation force correction for inviscid spheres in standing waves}}.  \jt{J. Acoust. Soc. Am.}  \bvol{142}~(3),  \pg{1167--1170}.

\bibitem[Naka \& Hasegawa(2020)]{Naka2020Dec}
{\sc \au{Naka, M.} \& \au{Hasegawa, K.}} \yr{2020}  \at{{Breakup characteristics of levitated droplets in a resonant acoustic field}}.  \jt{Phys. Fluids}  \bvol{32}~(12).

\bibitem[Pazos~Ospina {\em et~al.\/}(2022)Pazos~Ospina, Contreras, Estrada-Morales, Baresch, Ealo \& Volke-Sep{\ifmmode\acute{u}\else\'{u}\fi}lveda]{ospina2022particle}
{\sc \au{Pazos~Ospina, Jhon~F.}, \au{Contreras, Victor}, \au{Estrada-Morales, Jordan}, \au{Baresch, Diego}, \au{Ealo, Joao~Luis} \& \au{Volke-Sep{\ifmmode\acute{u}\else\'{u}\fi}lveda, Karen}} \yr{2022}  \at{{Particle-Size Effect in Airborne Standing-Wave Acoustic Levitation: Trapping Particles at Pressure Antinodes}}.  \jt{Phys. Rev. Appl.}  \bvol{18}~(3),  \pg{034026}.

\bibitem[Petersson {\em et~al.\/}(2007)Petersson, {\AA}berg, Sw{\ifmmode\ddot{a}\else\"{a}\fi}rd-Nilsson \& Laurell]{petersson2007free}
{\sc \au{Petersson, Filip}, \au{{\AA}berg, Lena}, \au{Sw{\ifmmode\ddot{a}\else\"{a}\fi}rd-Nilsson, Ann-Margret} \& \au{Laurell, Thomas}} \yr{2007}  \at{{Free Flow Acoustophoresis:{\hspace{0.167em}} Microfluidic-Based Mode of Particle and Cell Separation}}.  \jt{Anal. Chem.}  \bvol{79}~(14),  \pg{5117--5123}.

\bibitem[Petersson {\em et~al.\/}(2005)Petersson, Nilsson, Holm, J{\ifmmode\ddot{o}\else\"{o}\fi}nsson \& Laurell]{petersson2005continuous}
{\sc \au{Petersson, Filip}, \au{Nilsson, Andreas}, \au{Holm, Cecilia}, \au{J{\ifmmode\ddot{o}\else\"{o}\fi}nsson, Henrik} \& \au{Laurell, Thomas}} \yr{2005}  \at{{Continuous separation of lipid particles from erythrocytes by means of laminar flow and acoustic standing wave forces}}.  \jt{Lab Chip}  \bvol{5}~(1),  \pg{20--22}.

\bibitem[Rajendran {\em et~al.\/}(2022)Rajendran, Jayakumar, Azharudeen \& Subramani]{Rajendran2022Jun}
{\sc \au{Rajendran, Varun~Kumar}, \au{Jayakumar, Sujith}, \au{Azharudeen, Mohammed} \& \au{Subramani, Karthick}} \yr{2022}  \at{{Theory of nonlinear acoustic forces acting on inhomogeneous fluids}}.  \jt{J. Fluid Mech.}  \bvol{940}.

\bibitem[Rueckner {\em et~al.\/}(2023)Rueckner, Peidle, Crockett \& Davis]{Rueckner2023Aug}
{\sc \au{Rueckner, Wolfgang}, \au{Peidle, Joseph}, \au{Crockett, Allen} \& \au{Davis, Daniel}} \yr{2023}  \at{{Particle size effects on stable levitation positions in acoustic standing waves}}.  \jt{J. Acoust. Soc. Am.}  \bvol{154}~(2),  \pg{1339--1346},  \arxiv{arXiv: 37650782}.

\bibitem[Shi \& Apfel(1996)]{Shi1996Apr}
{\sc \au{Shi, W.~Tao} \& \au{Apfel, Robert~E.}} \yr{1996}  \at{{Deformation and position of acoustically levitated liquid drops}}.  \jt{J. Acoust. Soc. Am.}  \bvol{99}~(4),  \pg{1977--1984}.

\bibitem[Stone \& Leal(1989)]{Stone1989Jan}
{\sc \au{Stone, H.~A.} \& \au{Leal, L.~G.}} \yr{1989}  \at{{Relaxation and breakup of an initially extended drop in an otherwise quiescent fluid}}.  \jt{J. Fluid Mech.}  \bvol{198},  \pg{399--427}.

\bibitem[Thirisangu {\em et~al.\/}(2023)Thirisangu, Hemachandran \& Subramani]{Thirisangu2023Dec}
{\sc \au{Thirisangu, Jeyapradhap}, \au{Hemachandran, E.} \& \au{Subramani, Karthick}} \yr{2023}  \at{{Suspending droplets beyond the Rayleigh limit: The interplay of acoustic and gravity forces}}.  \jt{Phys. Fluids}  \bvol{35}~(12),  \pg{122012}.

\bibitem[Thirisangu {\em et~al.\/}(2024)Thirisangu, Rajendran, Selvakannan, Jayakumar, Hemachandran \& Subramani]{Thirisangu2024Aug}
{\sc \au{Thirisangu, Jeyapradhap}, \au{Rajendran, Varun~Kumar}, \au{Selvakannan, Snekan}, \au{Jayakumar, Sujith}, \au{Hemachandran, E.} \& \au{Subramani, Karthick}} \yr{2024}  \at{{Droplets in Acoustic Fields: A Unified Theory from Migration to Splitting}}.  \jt{arXiv} ,  \arxiv{arXiv: 2408.06092}.

\bibitem[Tian {\em et~al.\/}(1993)Tian, Holt \& Apfel]{Tian1993Jun}
{\sc \au{Tian, Yuren}, \au{Holt, R.~Glynn} \& \au{Apfel, Robert~E.}} \yr{1993}  \at{{Deformation and location of an acoustically levitated liquid drop}}.  \jt{J. Acoust. Soc. Am.}  \bvol{93}~(6),  \pg{3096--3104}.

\bibitem[Yarin {\em et~al.\/}(1998)Yarin, Pfaffenlehner \& Tropea]{Yarin1998Feb}
{\sc \au{Yarin, A.~L.}, \au{Pfaffenlehner, M.} \& \au{Tropea, C.}} \yr{1998}  \at{{On the acoustic levitation of droplets}}.  \jt{J. Fluid Mech.}  \bvol{356},  \pg{65--91}.

\bibitem[Yosioka \& Kawasima(1955)]{yosioka1955acoustic}
{\sc \au{Yosioka, Kawashima} \& \au{Kawasima, Yukihiko}} \yr{1955}  \at{Acoustic radiation pressure on a compressible sphere}.  \jt{Acta Acustica united with Acustica}  \bvol{5}~(3),  \pg{167--173}.

\bibitem[Zang {\em et~al.\/}(2017)Zang, Lin, Li, Chen, Li \& Geng]{Zang2017Mar}
{\sc \au{Zang, Duyang}, \au{Lin, Kejun}, \au{Li, Lin}, \au{Chen, Zhen}, \au{Li, Xiaoguang} \& \au{Geng, Xingguo}} \yr{2017}  \at{{Acoustic levitation of soap bubbles in air: Beyond the half-wavelength limit of sound}}.  \jt{Appl. Phys. Lett.}  \bvol{110}~(12),  \pg{121602}.

\end{thebibliography}
\end{document}